# A new concept of (open) TE$_{011}$ cylindrical cavity

G. Annino[@], M. Cassettari, and M. Martinelli
*Istituto per i Processi Chimico-Fisici, Area della Ricerca CNR, via G. Moruzzi 1, 56124 Pisa (Italy)*
(Dated: December 6, 2006)

**Abstract**

The confinement properties of the open structure formed crossing a circular waveguide perpendicular to a parallel-plate waveguide are discussed, highlighting the fundamental differences with respect to the common high-frequency resonators. Among the electromagnetic modes trapped at the intersection region of the two waveguides, the TE$_{011}$ one appears as the most appropriate for high-frequency applications. The experimental characterization of this mode is described in detail, investigating the response of a millimeter wave configuration resonating at 281 GHz, which shows state-of-the-art performances. The properties of the TE$_{011}$ mode are studied in terms of the geometry, calculating the mode chart and the related quality factor and power-to-field conversion efficiency. The mode chart is then determined for configurations including a sample holder, in which one of the component waveguides is filled with a low-loss dielectric material. The TE$_{011}$ mode reveals in general remarkable merit figures, as well as a significant stability with respect to the geometrical imperfections and to the insertion of a sample holder. The obtained results show that the proposed single-mode resonator competes with the standard cavities in terms of performance, versatility, and simplicity.

---

[@] *e-mail address: geannino@ipcf.cnr.it*

## 1. Introduction

The millimeter wave region of the electromagnetic spectrum knew in the last years a rapidly growing activity, following the improved accessibility to these wavelengths guaranteed by the development in source and detector technology. A relevant effort was in particular dedicated to the realization of the basic passive components of a typical spectroscopic setup, as low-loss waveguides [1], polarizers [1, 2], non-reciprocal elements [3], and resonators [4-6]. In the last case, the solutions commonly adopted at millimeter wavelengths are borrowed from the adjacent spectral regions, where the technology of the resonant cavities is well established. The main challenge in the development of resonators is the realization of a device combining the performances of the single-mode cavities, typically employed at microwaves, with the versatility of the overmoded Fabry-Perot resonators, mainly employed at far infrared and higher frequencies. On the other hand, such device should not present the counterpart of the above benefits, namely the mechanical complexity of single-mode cavities and the extended volume of Fabry-Perot, which often represent a limitation for millimeter wave applications.

It has been recently demonstrated that a system of coupled waveguides shows a variety of electromagnetic modes trapped at their intersection region [7-17]. Such system can behave as a single-mode resonator which combines a very simplified spectrum to high performances, according to the preliminary analysis of Ref. 18. Among the proposed geometries, the intersection of a cylindrical waveguide with a parallel-plate waveguide appears as the most appropriate for practical applications, in virtue of its largely open structure and to the presence of transverse electric (TE) modes. The trapping of electromagnetic radiation at the intersection of the waveguides is in general due to a proper combination of symmetry of the configuration and cutoff frequency of the modes propagating along the waveguides. Such mechanism of confinement is rather dissimilar from that of a close metallic box resonator, based on a complete shielding of the radiation, as well as from that of the Fabry-Perot, explainable in terms of a simple geometrical optics representation. The confinement of electromagnetic radiation at the intersection of coupled waveguides lead therefore to a new conception of open resonator, the design and the operation of which requires a dedicated approach.

The aim of this article is a systematic investigation of the properties of the $TE_{011}$ mode trapped at the intersection between a cylindrical waveguide and a parallel-plate waveguide. Such investigation will first consider the experimental arrangement enabling the observation and the characterization of the mode at millimeter wavelenghts. The intrinsic parameters of the mode will be studied through a numerical modelling, which will determine the mode chart, the quality factor and the power-to-field conversion efficiency obtainable in the allowed configurations. The mode chart will then be calculated for practical working configurations in which a dielectric sample holder is inserted in the resonator. The obtained experimental and theoretical results will show that the proposed open single-mode resonator represents a favourable synthesis of ease of realization, open structure, and high performances, which prefigures its employment in demanding applications.

## 2. Theoretical background

The demonstration of the existence of electromagnetic (e.m.) modes trapped at the intersection region of coupled metallic waveguides represents a relatively recent achievement. The first results, based on the TE modes of crossed rectangular waveguides [13-17], have been generalized to the hybrid modes of configurations with different waveguides [18], and to configurations including dielectric regions [19].

The common aspect of such systems is the existence of confined solutions with frequency below and above the cutoff of the modes propagating along the waveguides. The basic reason that ensures the confinement of the radiation is the availability, at the intersection region, of a volume with minimum linear size larger than that available in each individual waveguide [11]. In virtue of the enlarged volume, some electromagnetic modes can exist at frequencies lower than the lowest cutoff of the propagating modes. Such isolated modes are necessarily confined because, in the limit of infinite waveguides, they do not have any decay channel. On the other hand, some confined solutions can exist above the lowest cutoff of the crossed waveguides. Such modes are embedded in the continuous spectrum of propagating modes, which therefore represent possible leakage channels. The additional mechanism that forbids the radiation leakage is in general given by the symmetry of the configuration, which introduces selection rules for the coupling of the confined modes to the propagating ones. It follows, accordingly, that the symmetry plays a basic role in the definition of the confinement properties of coupled waveguides. On the basis of its analysis, for instance, the minimum frequency interval in which a mode with a given symmetry can be trapped is easily predictable [18].

In the idealized condition of perfect conductors, exact geometry and infinite waveguides, the trapped modes appear as resonances with infinite quality factor. In any practical case, the imperfect geometry and the finite conductivity and length of the employed waveguides limit the figure of merit of the resonant device. Nevertheless, the quality factor achievable in the presence of the ohmic losses alone can be quite high, also at millimeter wavelengths. Analogously, the violation of the symmetry due to the geometrical imperfections introduces in general marginal losses at the level of fabrication tolerances which is typical of these wavelengths [18]. The unloaded merit factor $Q_0$ of such a resonator can be therefore quite high. Its excitation requires a coupling between the electromagnetic energy stored in the intersection region and that propagating in external guiding systems. The coupling can be obtained employing waveguides with finite length. Such operation introduces, as usual, a certain level of irradiation losses. However, being the excitation enhanced by the resonance effect, the coupling level necessary to reach a complete transfer of energy to the high $Q_0$ resonances can be quite low. As a consequence, the excitation setup is expected to introduce a weak perturbation on the field distribution of these modes.

In one of its simplest form, a coupled-waveguide resonator is realized intersecting a circular waveguide with a parallel-plate waveguide having parallel axis of symmetry [18, 19]. The resulting structure is shown in Fig. 1. When compared with the common millimeter wave resonators, such a structure shows a rather surprising topology. Contrarily to the case of close metallic cavities, the radiation stored in the intersection region can escape both radially and axially, through large apertures which minimum size is comparable with that of the confinement region. The proposed resonator can be seen as a cylindrical metallic cavity in which the plungers are removed and the body (the remaining cylindrical waveguide) is cut in two parts. It is worth emphasizing that the cut in the body of the cavity is essential to obtain the trapping of some modes, since the simple cavity without plungers cannot basically resonate. The additional aperture in the open cylinder ensures the confinement of the radiation, instead to introduce a further leakage, as suggested by the common sense. On the other hand, such resonator is still more dissimilar to the open Fabry-Perot cavity. In the case of the Fabry-Perot, indeed, the confinement of the radiation is guaranteed by the concavity of the (real or fictitious) mirrors, which refocuses the radiation on each reflection on them. Here, the metallic surface is neither simply concave nor convex. The profile of such surface - with its sharp edges directed towards the storing region - would suggest the escape of the radiation, instead of its confinement. Nevertheless, it is possible to conclude that, in a sense, the edges help in taking in place the radiation [9]. The apparent paradox holds only in the geometrical optics representation, which is not applicable here due to the closeness between the size of the confinement region and the employed wavelength. A correct explanation of the confinement mechanism requires in fact a pure wave analysis.

The proposed resonator cannot be classified, therefore, in none of the common classes of resonant structures. Looking at the possible roots of this device, an evident analogy can be established with the nonradiative dielectric resonator discussed in Ref. 19, in the limit of unitary permittivity of the dielectric waveguide. There is however a fundamental aspect that differences the analysis developed in this work from that of Ref. 19. In the latter, the confinement of the radiation in the intersection region between the two waveguides is largely attributed to the dielectric component, following an inductive approach based on a continuous deformation of the structure and on its perturbative analysis. On the basis of this approach, it is not possible to predict the existence of modes confined at the intersection of empty metallic waveguides. The open resonator here proposed can be equally seen as a generalization of the open nonradiative cavity studied in Ref. 20, in which the two plungers are removed.

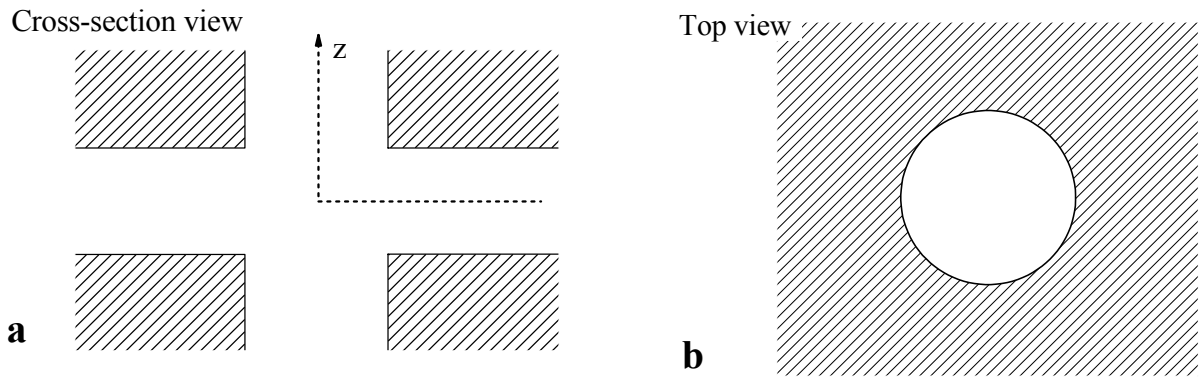

Fig. 1. *Sketch of the resonant device obtained coupling a circular waveguide to a parallel-plate waveguide, together with the excitation setup.* ***a****) Cross-section view. The dashed lines delimit the quarter of the axial cross section on which the e.m. field are calculated (see text).* ***b****) Top view.*

The spectrum of the coupled-waveguide resonators is composed by few resonances determined by the geometry of the configuration, in which the field distribution typically extends over about a wavelength [18]. The structures with rotational invariance show some peculiar characteristics. They are in general compatible with transverse electric (TE) and transverse magnetic (TM) modes with null azimuthal index (vanishing derivative along the azimuthal coordinate), as demonstrated in Appendix on the basis of a general approach. A confined $TE_{011}$ mode has been observed in various open structures with such symmetry [18, 19, 21]. The research presented in the following will focus on this mode, which is of remarkable importance in practical applications [4, 22-27].

### 3. Experimental characterization

The coupling of e.m. energy from a propagating system to the mode confined in the resonator requires an overlap between the field distributions of the two structures. As anticipated, the resonant mode can be made accessible limiting the extension of one or both of the intersecting waveguides. In this manner, a fraction of the energy of the mode lies outside the resonant structure. A possible implementation of such an excitation scheme is shown in Fig. 2. The right part of the figure reports a top view of the resonator, in which the cylindrical waveguide is drilled close to a side of the parallel-plate waveguide. A proper design of this geometry allows the introduction of a controlled radiation leakage from the resonance mode. Such radiation can be coupled to the propagating system, placing for instance an external waveguide close to the resonant assembly. The level of the coupling, and then the amount of energy transferred to the resonator, can be adjusted changing the distance between the waveguide and the active region of the resonator. For a practical convenience, the external waveguide should be in contact with the resonator. Such solution has the advantage of a high mechanical stability and of a reduced leakage of the incoming radiation. The highest efficiency in the excitation of the $TE_{011}$ mode is obtained for radiation polarized along the parallel plates, following the field distributions expected for this mode. The level of the coupling can be finally controlled moving the resonator orthogonally to its axis, as indicated in the right part of Fig. 2.

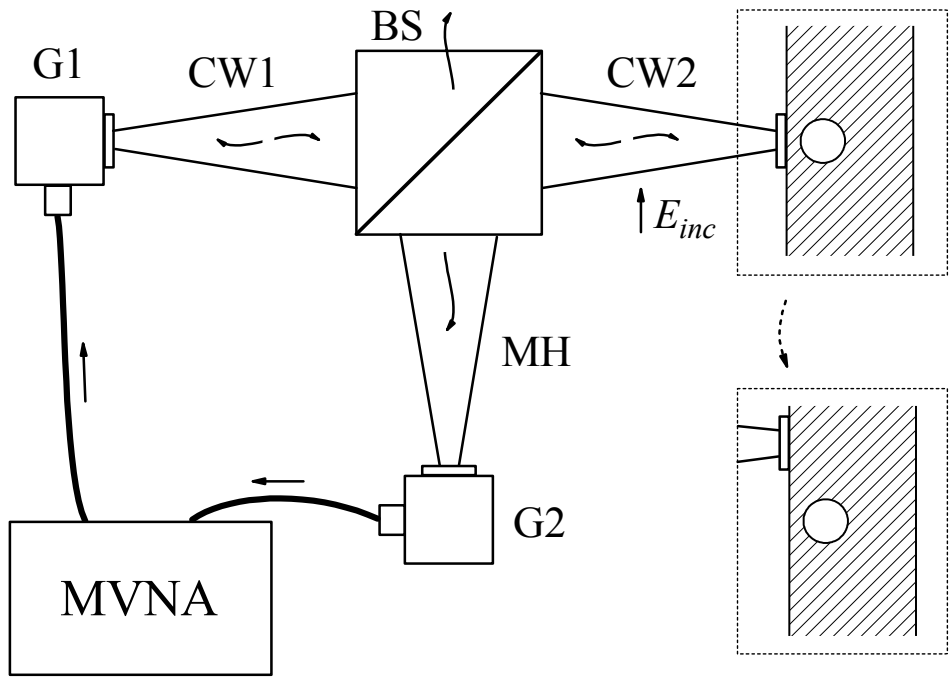

Fig. 2. *Schematic representation of the experimental setup. MVNA indicates the vector network analyzer, G1 the Gunn diode acting as a source, CW1 the corrugated transition in the source arm, BS the beam splitter, CW2 the corrugated transition in the resonator arm, MH the metallic horn, G2 the Gunn diode acting as local oscillator, and $E_{inc}$ the microwave electric field, directed along the parallel plates. The undulated arrows represent the propagating waves. The coupling level is adjusted moving the resonant assembly as illustrated by the two configurations in the dashed box.*

The remaining part of the figure illustrates the measurement setup employed in the present work. The generation of the radiation, and its detection, was based on a millimeter wave vector network analyzer (MVNA) in the so-called ESA2 configuration, which covers almost continuously the interval 250-800 GHz (ABmm, Paris) [28]. The millimeter wavelength is generated mixing the signal emitted by an Yttrium-Iron-Garnet source in the 8-18 GHz interval with that emitted by 2 Gunn diodes. The first Gunn, covering the frequency interval from 78 to 112 GHz allows the upconversion of the microwave frequency. The second Gunn, covering the interval 82-103 GHz, is employed as local oscillator in the detection arm. The final frequency is generated through a further multiplication stage, employing a nonlinear element (Schottky diode). The dynamic range obtained with such configuration is of the order of 130 dB at 285 GHz. The emitted radiation propagates along a corrugated horn designed to work around 285 GHz. The horn ensures a smooth transition from a rectangular waveguide with cross section of 0.71 mm x 0.355 mm, to a circular waveguide with 18 mm diameter. The radiation is then divided in two parts by means of a thin Mylar foil acting as beam splitter. The wave propagating through the beam splitter is coupled to another corrugated transition that, starting from a circular cross-section with 18 mm diameter, focuses the radiation on a circular cross-section with 2 mm diameter. The end of such transition is put in contact with the resonator, following the scheme of Fig. 2. The radiation reflected by the resonator propagates back to the beam splitter, where it is partially decoupled from the incoming wave and redirected towards the detector through a metallic horn.

At millimeter wavelengths, the standing waves generated along the propagation circuit and the power variations of the source lead to a power level on the detector rapidly fluctuating with the frequency. This variation leads in turn to a distortion of the resonance curve. The intrinsic shape of the resonance can be largely recovered through a normalization procedure, consisting in the division of the signal obtained with the excitation waveguide close to the resonator to that obtained with the waveguide far enough from it, according to the arrangements of Fig. 2.

The setup and the procedure discussed above have been applied to the characterization of a coupled-waveguide resonator composed by two square copper plates with 18 mm thickness and 33 mm side, kept at a distance of $0.43 \pm 0.01$ mm. The nominal diameter of the hole representing the circular waveguide was $1.250 \pm 0.001$ mm, and the distance between the hole and the closest lateral edge $0.2 \pm 0.02$ mm. The flat surfaces of the structure were optically polished. The surface of the hole was grinded by using a steel sphere with $1.250 \pm 0.001$ mm diameter.

The theoretical spectrum of such structure was calculated by means of the finite-element software Multiphysics 3.2a (COMSOL, Sweden) neglecting the presence of the excitation setup, namely in a simplified geometry similar to that shown in Fig. 1. According to the numerical modelling, the above resonator supports a trapped $TE_{011}$ mode resonating at $280.07 \pm 1.3$ GHz, where the uncertainty in the calculated frequency is determined by the fabrication tolerances.

The frequency interval around the expected resonance frequency was investigated comparing the signal reflected by the resonator for different positions of the excitation waveguide. An absorption peak was clearly visible close to the predicted frequency. In particular, Fig. 3 reports the signal obtained with the waveguide centered on the resonance region with that obtained with the waveguide completely decoupled from the resonator. The normalization of the resonance curve yields the spectrum of the device, shown in Fig. 4. The weak oscillation outside the central peak represents a residual effect of the variation of the power in the excitation circuit.

The resonance frequency of the observed mode is given by $v_0$=281.24 GHz, which agrees quite well with the predicted value, taking into account the fabrication tolerances. Such agreement confirms the weak perturbation introduced by the excitation setup. A similar result follows from the numerical analysis of the complete working configuration. The loaded quality factor of the mode, $Q_L$, was obtained by means of a lorentzian fit of the resonance curve, calculated in the complex plane representing the amplitude and the phase of the signal, exploiting the vector capabilities of the millimeter wave network analyzer. The result of the fit is shown in the left inset of Fig. 4. The polar plot shows in particular that the resonance is slightly overcoupled, since the origin of the plot lies inside the resonance curve, and characterized by $Q_L = 1000$. This quantity is partially determined by the intrinsic losses in the resonator, and partially by the level of the coupling. The intrinsic losses can be expressed in terms of the unloaded quality factor $Q_0$. Such figure is related to $Q_L$ and to the power reflected at the resonance $P_r$, normalized to the power $P_0$ reflected without coupling to the resonator. The relation involving these quantities is given by $Q_L = \dfrac{Q_0}{2}\left(1 \pm \sqrt{\dfrac{P_r}{P_0}}\right)$ [29, 30]. In such expression, the positive sign applies to the undercoupled conditions and the negative sign to the overcoupled ones. Despite the simplicity of the structure, the unloaded merit factor of the $TE_{011}$ mode, calculated by means of the above formula, exhibits the remarkable value of $Q_0 = 2100$ at 281 GHz, which represents the state of the art for single-mode metallic cavities at such frequencies. Such value can be compared with that imposed by the copper conductivity, indicated as $Q_{0,\Omega}$. It results, in particular, that $Q_{0,\Omega} = 4100$ for the investigated geometry, as discussed in the following. The difference between the observed and the calculated values of this quantity can be ascribed to the level of finishing of the surfaces and to a residual asymmetry in the actual geometry of the resonator [18].

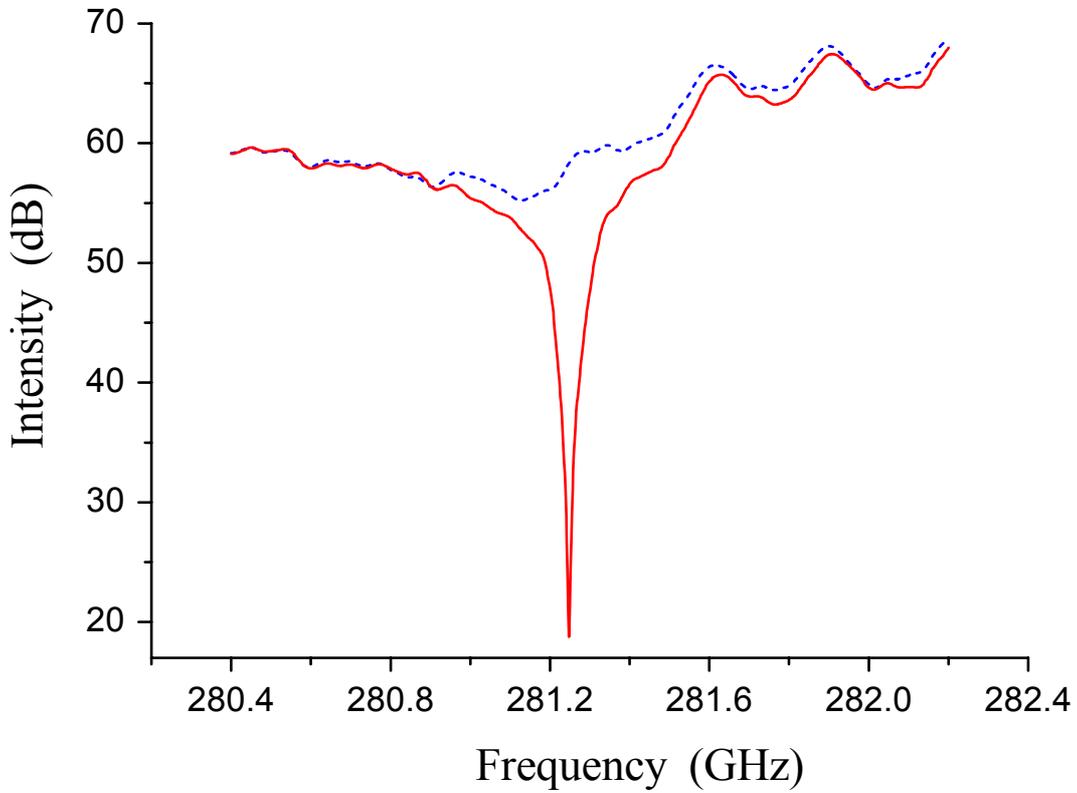

Fig. 3. *Experimental curves obtained for different coupling conditions. Solid line: excitation waveguide centered with respect to the axis of the resonator. Dashed line: excitation waveguide completely decoupled from the resonator.*

The electric field distribution of the observed mode on the axial cross section of the cavity is shown in the right inset of Fig. 4. Such distribution is plotted only on a quarter of the cross section, in virtue of the azimuthal invariance of the mode and of the mirror symmetry of the structure with respect to the median plane of the parallel plates. The condition of radial confinement of the $TE_{011}$ mode follows directly from its field structure. As shown in Appendix, indeed, it does not share any field component with the cutoff-less transverse electromagnetic modes (TEM) propagating along the parallel-plate waveguide. Accordingly, the radial leakage vanishes provided that $l < \frac{\lambda_0}{2}$, being in turn $l = \frac{\lambda_0}{2}$ the first cutoff condition of the parallel-plate TE modes. Here $\lambda_0$ is the resonant free-space wavelength.

Analogously, the use of an electromagnetic wave linearly polarized along the parallel plates leads, outside the resonance frequency, to an almost complete reflection at the interface between the waveguide and the resonator, until the condition $l < \frac{\lambda_{0,rad}}{2}$ is satisfied; in this context, $\lambda_{0,rad}$ represents the wavelength of the incoming radiation. Under the above condition, the employed excitation configuration corresponds to a classical reflection scheme.

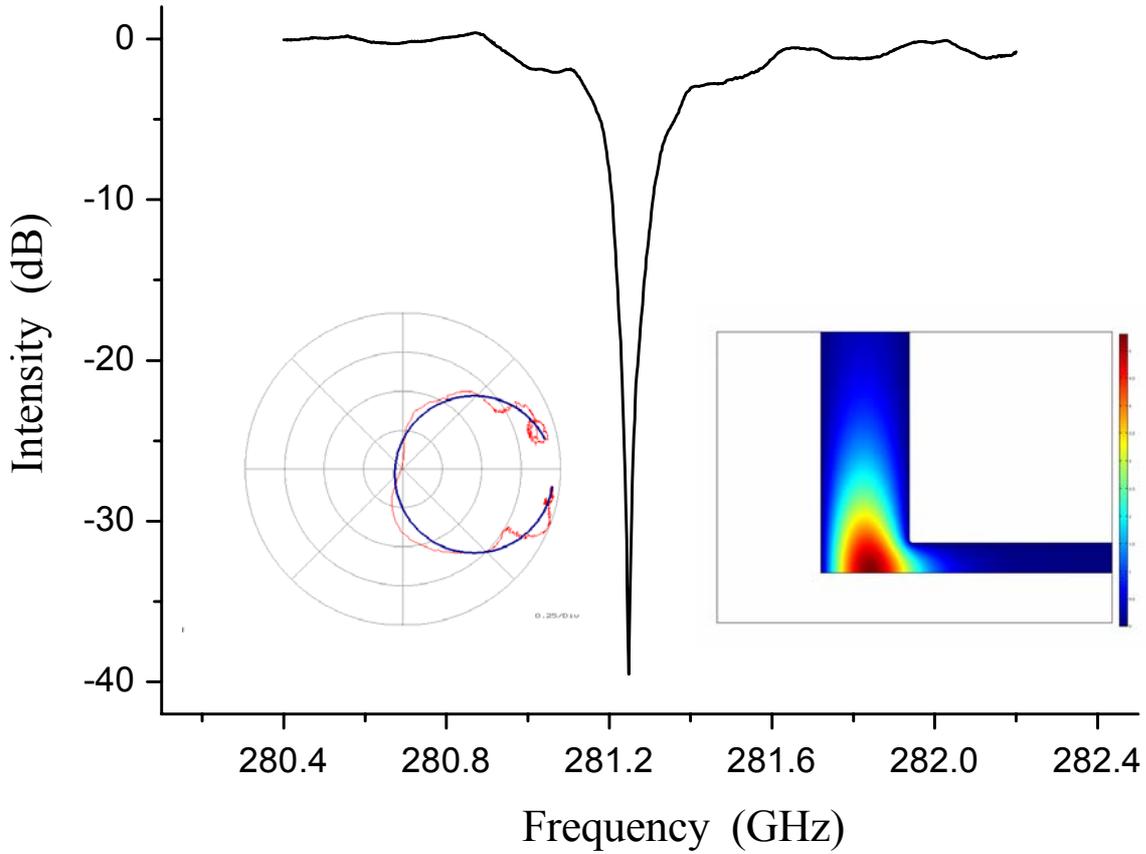

Fig. 4. *Normalized absorption spectrum of the $TE_{011}$ mode. Right inset: calculated electric field distribution on a quarter of the axial cross section of the resonator. On the right, the colors (gray) scale, expressed in arbitrary units. Left inset: polar plot of the normalized absorption spectrum, together with the lorentzian fit (thick-line circle).*

The axial confinement of the $TE_{011}$ mode is ensured when $\lambda_0$ satisfies the condition $\frac{r}{\lambda_0} < \frac{3.8317}{2\pi}$, where $\frac{r}{\lambda_0} = \frac{3.8317}{2\pi}$ is the first cutoff of the $TE_0$ modes propagating along the circular waveguide [19]. An alternative way to excite the mode can be realized coupling the radiation through the circular waveguide, as experimentally verified. In this case, however, the complete reflection on the resonator can be frustrated by the coupling with the $TE_{11}$ mode propagating along the cylindrical waveguide.

## 4. Mode chart

A complete information on the possible geometry of a resonant device, and on its relation with the frequency of a specific mode, can be encoded in a single graph, namely in the mode chart of the device. The mode chart of the $TE_{011}$ mode is reported in Fig. 5, in terms of the radius $r$ of the cylindrical waveguide and of the distance $l$ between the parallel plates, both normalized to $\lambda_0$. In order to avoid the divergence of the e.m. field at the sharp edges delimiting the intersection region, the edges were replaced by a circular profile with radius of curvature $\frac{\rho}{\lambda_0} = 25 \cdot 10^{-3}$.

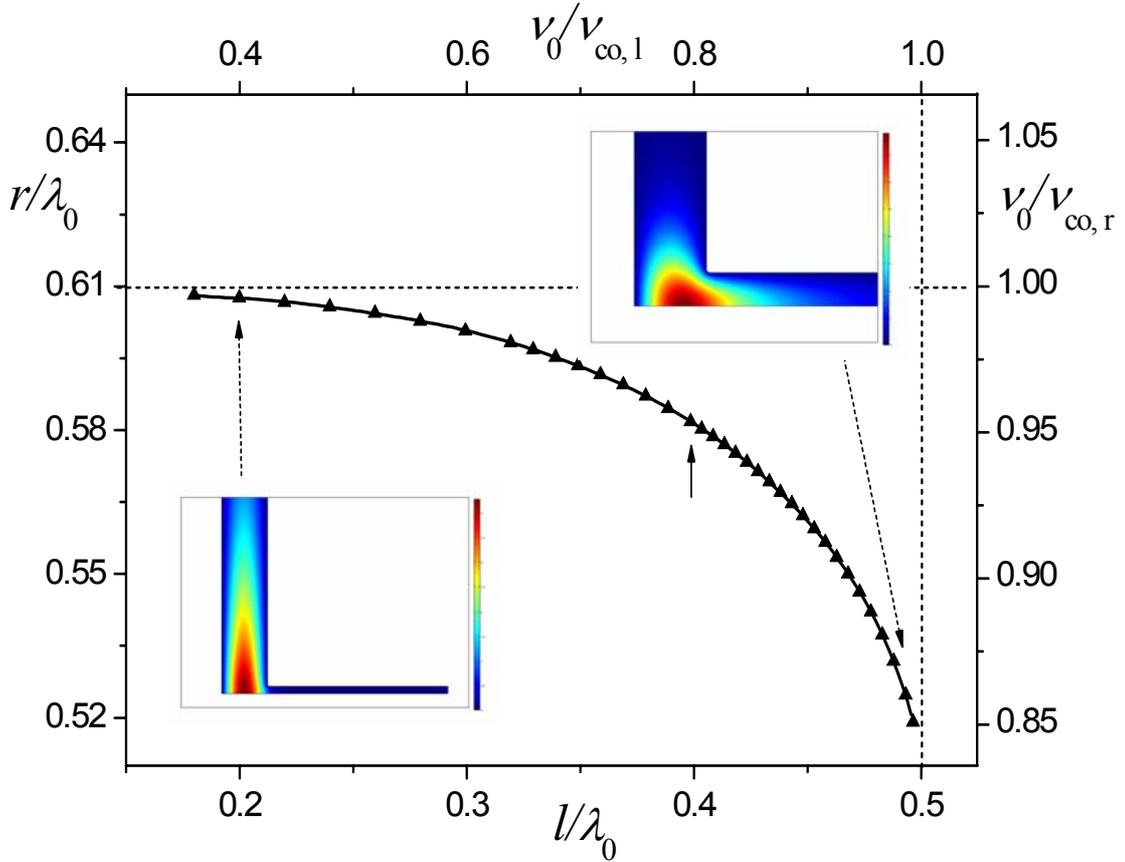

Fig. 5. *Mode chart of the $TE_{011}$ mode (see text for the meaning of the labels). The up triangles represent the calculated values, the solid line the fitting curve. The horizontal and vertical dashed lines indicate the theoretical asymptotes. Insets: electric field distribution corresponding to the geometry indicated by the dashed arrows. On the right, the colors (gray) scale, expressed in arbitrary units. The solid arrow indicates the configuration considered in Fig. 3.*

The insets of Fig. 5 show the field distribution of the $TE_{011}$ mode, calculated for the limit conditions indicated by the two external dashed arrows. The solid arrow in the middle of the curve indicates the configuration to which refers the field distribution of Fig. 4. The functional relationship $\frac{r}{\lambda_0}\left(\frac{l}{\lambda_0}\right)$ inside the interval of modelling, $\frac{l}{\lambda_0} \in (0.18, 0.497)$, can be reproduced very accurately by a 9$^{th}$ degree polynomial curve $\frac{r}{\lambda_0} = \sum_{j=0}^{9} a_j \left(\frac{l}{\lambda_0}\right)^j$. The values of the coefficients $a_j$ are reported in the Table. Such polynomial function generates the fitting curve of Fig. 5. The relationship between $r$ and $l$ is monotonic; such property allows obtaining the resonance frequency of an arbitrary geometry.

| | |
|---|---|
| $a_0$ | 17.64 |
| $a_1$ | -522.62 |
| $a_2$ | 7021 |
| $a_3$ | -54208.7 |
| $a_4$ | 265156 |
| $a_5$ | -852422 |
| $a_6$ | 1.80184E6 |
| $a_7$ | -2.4161E6 |
| $a_8$ | 1.86592E6 |
| $a_9$ | -632737 |

Table. *Coefficients of the 9$^{th}$ degree polynomial function fitting the mode chart curve. The standard deviation of the fit is 1.06\*10$^{-4}$.*

In the central part of the $\frac{r}{\lambda_0}$ vs. $\frac{l}{\lambda_0}$ curve, the electromagnetic energy outside the intersection region is almost equally distributed between the cylindrical waveguide and the parallel-plate waveguide. Moving away from this condition, the curve shows an asymptotic convergence in both directions, towards the points $\frac{l}{\lambda_0} = a_1$ and $\frac{r}{\lambda_0} = a_2$. The two asymptotes $a_1$ and $a_2$ correspond very well with the theoretical values $a_1 = \frac{1}{2}$ and $a_2 = \frac{3.8317}{2\pi}$, indicated in Fig. 5 by the dashed lines. In the limit in which one of the two variables $\frac{r}{\lambda_0}$ and $\frac{l}{\lambda_0}$ tends to its asymptotic value, the other tends to zero. The field distribution becomes accordingly more and more extended along one waveguide, and only marginally extended in the other. Nevertheless, the narrow waveguide is still capable to induce a trapping of the radiation in the intersection region. This is true, in principle, for any finite size of such guide, at least in the limit of perfectly conducting walls [31, 32]. However, when the secondary guide is very narrow, the mode is so extended along the main one that it cannot be trapped in structures with reasonable size. In the case of Fig. 5, the calculation has been stopped when the relative difference between the variable and its expected asymptote was lower than 1%. The confinement level of the mode can also be evaluated comparing its resonance frequency with the cutoff frequencies of the intersecting waveguides. The relative value of such frequencies is indicated in the right coordinate axis and in the up abscissa axis of Fig. 5. Here, $\nu_0$ is the resonance frequency, $\nu_{co,l}$ the first cutoff of the TE modes of the parallel-plate waveguide, $\nu_{co,r}$ the first cutoff of the TE$_0$ modes of the circular waveguide.

According to the mode chart of Fig. 5, the TE$_{011}$ mode exists for a wide range of geometrical conditions. Its practical applicability is strictly related to the figure of merit of the resonance, which is usually expressed in terms of the unloaded quality factor $Q_0$ and of the power-to-field conversion efficiency. The latter figure gives the root mean square of the electric or magnetic field at a specific point of the resonator, for unitary dissipated power $P$. The quality factor is important in any application related to the capability of the device to discriminate in frequency, as in the case of frequency filters [33], or when the key parameter is the total amount of energy stored in the resonator, as for the laser sources [34]. The conversion factor is important when the figure of merit is related to the amplitude of the field inside the resonator. This is the case, for instance, of the absolute sensitivity in electron paramagnetic resonance, related to the magnetic induction field $\bar{B} = \sqrt{\langle B^2 \rangle}$ [35], or to the quantum electrodynamics applications, related to the electric field $\bar{E} = \sqrt{\langle E^2 \rangle}$ [36]. The brackets in the above expressions indicate the average over the oscillation period of the radiation. In the following, the emphasis will be put on the behaviour of the magnetic field $\bar{B}$; similar results were obtained, however, for the electric field.

## 5. Performances

For a copper cavity resonating at 300 GHz, the calculated quality factor $Q_0$ and axial conversion factor $\bar{B}_z$ (on the point of highest magnetic field) are shown in Fig. 6. The maximum values of these quantities are $Q_0 = 4930$ and $\bar{B}_z = 36.9$ G/W$^{1/2}$; the latter figure is obtained for $\frac{l}{\lambda_0} = 0.39$ and $\frac{r}{\lambda_0} = 0.587$, to which corresponds $Q_0 = 4080$. The resonator experimentally investigated corresponds to a configuration with $\frac{l}{\lambda_{0,\exp}} = 0.40$ and $\frac{r}{\lambda_{0,\exp}} = 0.583$, as indicated by the arrow in Fig. 5, which is very close to the condition of highest $\bar{B}_z$ conversion factor. The obtained quality factor $Q_{0,\exp} = 2100$ and conversion factor $\bar{B}_{z,\exp} = 25$ G/W$^{1/2}$ can be compared with the theoretical value expected for the investigated configuration, given by $Q_0 = 4100$ and $\bar{B}_z = 35$ G/W$^{1/2}$, respectively. These latter figures have been calculated taking into account their frequency dependence, discussed in the following.

The performances of the resonator depend on the radius of curvature of the profile which joins the intersecting waveguides. In particular, at $\frac{\rho}{\lambda_0} = 100 \cdot 10^{-3}$ the maximum quality factor and conversion factor are $Q_0 = 4900$ and $\bar{B}_z = 38.8$ G/W$^{1/2}$, the latter value being obtained for $Q_0 = 4410$ at $\frac{l}{\lambda_0} = 0.357$ and $\frac{r}{\lambda_0} = 0.579$. Such values can be compared with those of a close cylindrical cavity made with the same material and resonating at the same frequency. The TE$_{011}$ mode of a cylindrical cavity shows indeed a field distribution similar to that of the mode here investigated; it represents therefore a natural term of comparison for the expected performances. The quality factor and the conversion factor calculated for a copper cavity with optimal aspect ratio are $Q_{0,c}$ =5450 and $\bar{B}_{z,c}$ =46.3 G/W$^{1/2}$. In the practical applications, such ideal values are in general significantly reduced by the presence of the coupling hole and by that of the possible slits introduced in order to have an additional access to the cavity [4].

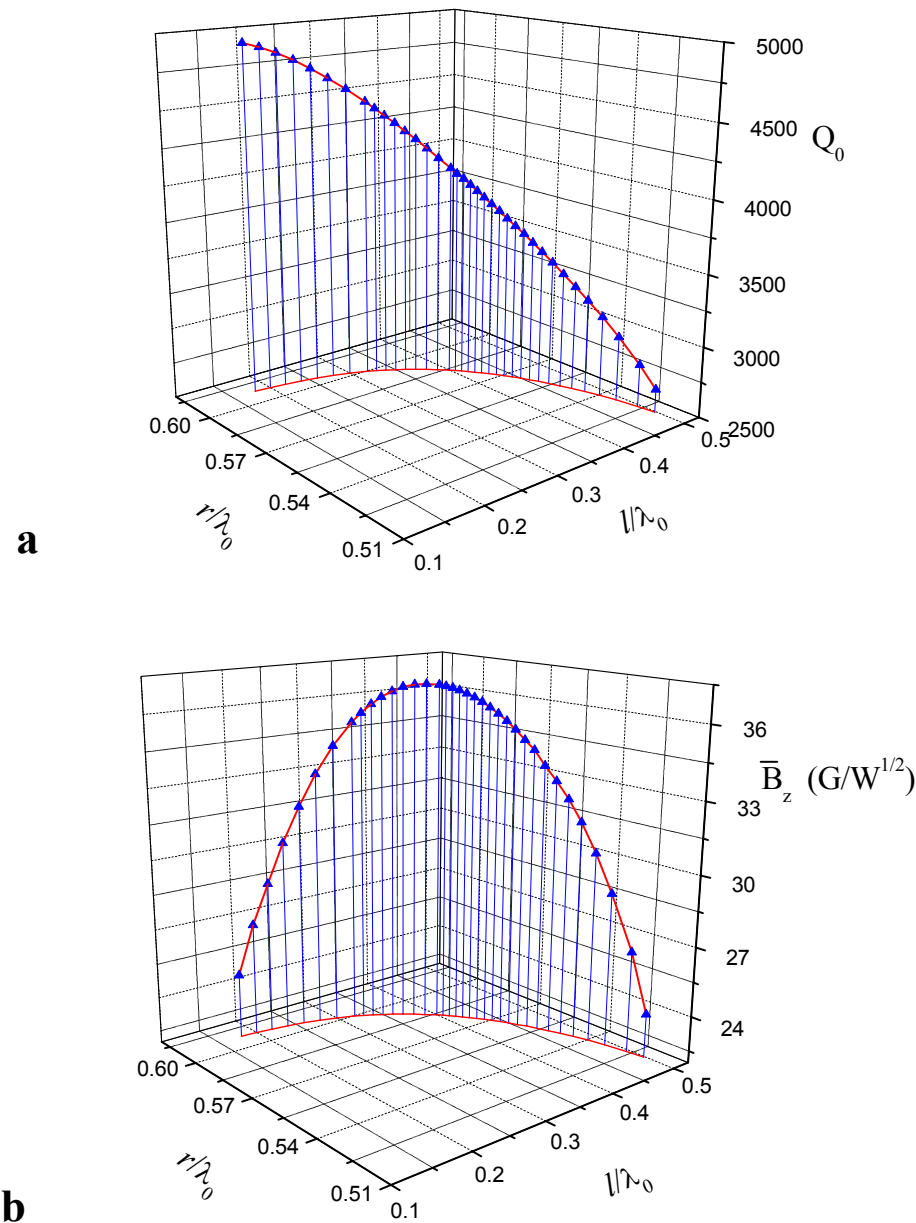

Fig. 6. *Calculated merit figures of a room temperature copper (resistivity 1.724 $\mu\Omega$ cm) resonator working at 300 GHz, as a function of the geometry. Up triangles: calculated points. Solid line: fitting curve. The solid line in the $rl$ plane represents the mode chart curve. a) Unloaded quality factor. b) Axial magnetic field conversion factor in the center of the resonator.*

In the absence of other sources of loss and assuming the resistivity of the conducting walls independent on the frequency $\omega$, $Q_0$ and $\bar{B}$ follow the simple scaling laws $Q_0 \propto \omega^{-1/2}$ and $\bar{B} \propto \omega^{3/4}$. Such relations can be deduced from the expressions $Q_0 = \dfrac{2 \int_{vol} H^2 dV}{\delta \int_{sur} H_t^2 dS}$ [37] and $\bar{B} = \sqrt{\langle B^2 \rangle} \propto \sqrt{\dfrac{W}{V}} = \sqrt{\dfrac{Q_0 \cdot P}{\omega_0 \cdot V}}$ [35]. Here $\delta$ is the skin depth in the conductor, $H$ the magnetic field, $H_t$ its component tangential to the conducting surface, W the total e.m. energy stored in the cavity, V its active volume [38], and $\omega_0$ the angular resonance frequency. Similar relations can be obtained for the dependence of the quality factor and of the conversion factor on the resistivity. In the limit of validity of the perturbative approach, in which the finite resistivity does not influence appreciably the e.m. field distribution, the scaling laws are given by $Q_0 \propto \rho^{-1/2}$ and $\bar{B} \propto \rho^{-1/4}$. Hence, the curves of Fig. 6 can be employed for any resistivity, geometry and frequency.

According to the above analysis, the performances of the proposed open resonator are competitive with those of a standard close cavity. The axial conversion factor $\bar{B}_z$ reaches its maximum value near the center of the mode chart, where the energy of the electromagnetic field shows the minimum spatial spreading. Close to the asymptotes, on the other hand, the active volume of the resonator is so large that the energy density goes to zero, as well as $\bar{B}$. On the contrary, the merit factor $Q_0$ shows a monotonic trend and its maximum value corresponds to the minimum allowed distance between the parallel metallic plates. This property suggests that a thin cut orthogonal to a long circular waveguide can trap a $TE_{011}$ mode with high Q value. In this case the electromagnetic energy is widely distributed along the circular waveguide, as shown in the inset of Fig. 5. Nevertheless, the ability of such configuration to discriminate in frequency can be remarkable. A small irregularity in straight waveguides can therefore introduce a sharp behaviour close to the cutoff, as anticipated.

## 6. Loaded resonator

In the common spectroscopic applications, it is necessary to introduce a sample in the cavity and very often also a sample holder. In view of a general use of the coupled-waveguide resonator, it is therefore mandatory to evaluate the effects of a low-loss sample holder on its mode chart. Two illustrative cases have been considered: in the first one, the cylindrical waveguide is filled with a dielectric tube made of teflon (dielectric permittivity $\varepsilon = 2.05$); in the second one, a teflon slab is introduced between the metallic plates. Both configurations can be easily realized, in virtue of the open structure of the resonator. The resulting mode charts of the $TE_{011}$ mode are reported in Fig. 7. The limit conditions for such configurations can be predicted following the same approach employed for the analysis of the empty resonator, taking now into account the effective optical size due to the dielectric material. In the case of cylindrical waveguide filled with a medium with permittivity $\varepsilon$, the asymptotes become $\dfrac{l}{\lambda_0} = \dfrac{1}{2}$, and $\dfrac{r}{\lambda_0} = \dfrac{3.8317}{2\pi\sqrt{\varepsilon}}$ [19]. When the parallel-plate waveguide is filled with a dielectric slab, the asymptotes are $\dfrac{l}{\lambda_0} = \dfrac{1}{2\sqrt{\varepsilon}}$, and $\dfrac{r}{\lambda_0} = \dfrac{3.8317}{2\pi}$. These limit conditions are well confirmed by the modelling. Similar results have been obtained for other representative cases covering a large interval of dielectric permittivity. The stability of the $TE_{011}$ resonance with respect to the insertion of a dielectric material along the intersecting waveguides makes the open cavity here considered particularly suited for millimeter wave dielectrometry applications.

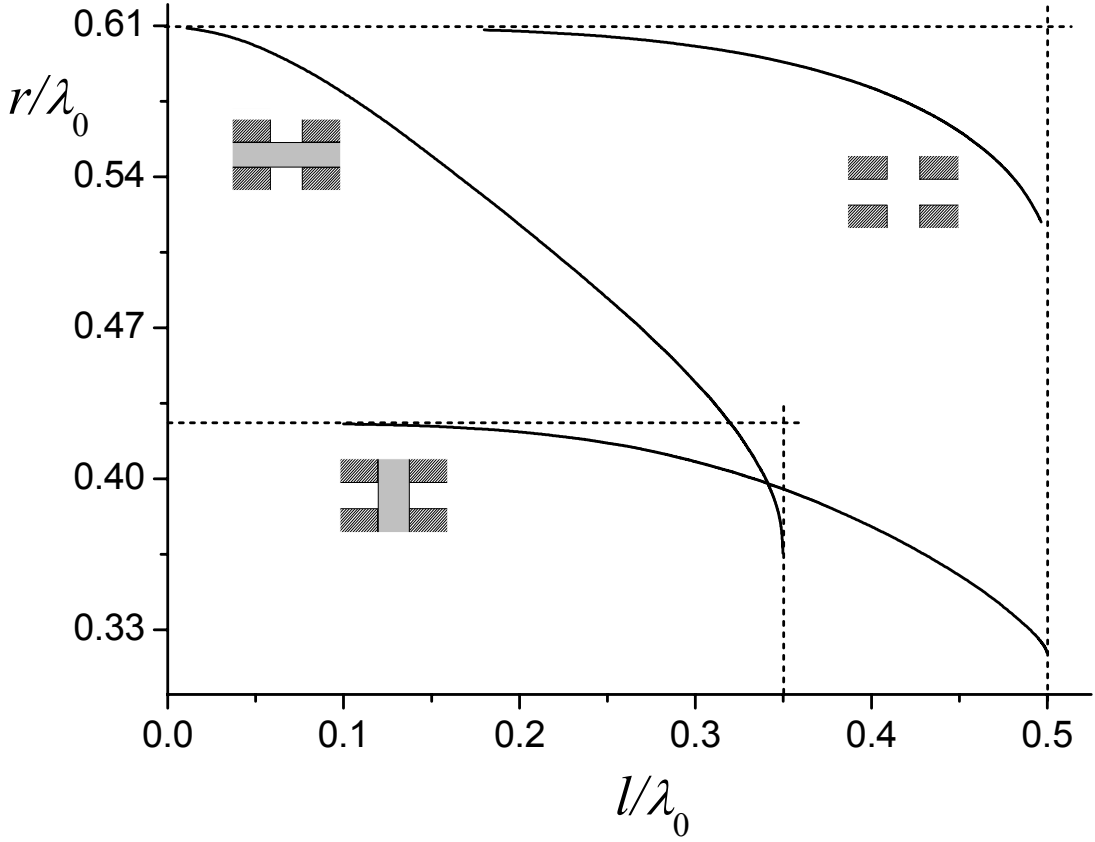

Fig. 7. *Mode chart of the TE$_{011}$ mode for different coupled-waveguide structures. The insets sketch the axial cross section of each of them. Top-right: empty resonator. Top-left: Teflon slab filling the parallel-plate waveguide. Bottom-left: Teflon tube filling the circular waveguide. The horizontal and vertical dashed lines indicate the theoretical asymptotes of the different configurations.*

From a practical point of view, another basic issue is the tuning of the resonance frequency. In the proposed resonator it can be implemented, for instance, varying the distance $l$ between the parallel plates. The dependence of the resonance frequency on $l$ for a given radius $r$ can be extracted from the mode chart of Fig. 5.

In conclusion, the resonator obtained coupling a circular waveguide to a parallel-plate waveguide shows different peculiarities. It combines a largely open structure to a reduced active volume, properties that represent the basis of its simplified spectrum of trapped modes. Among such modes, the TE$_{011}$ one ensures state-of-the-art performances at millimeter wavelengths and a weak dependence on the perturbation of the symmetry, without structural degeneracy with other confined modes. It is moreover stable with respect to the insertion of a sample holder along both the component waveguides. The capability of coupled waveguides to trap electromagnetic modes at their intersection region enables, therefore, the realization of a very simple and robust resonant device for high-frequency applications.

**Appendix**

The vector properties of the electromagnetic modes of a generic system with rotational invariance are considered first, assuming the azimuthal invariance ($\frac{\partial}{\partial \varphi} \equiv 0$) of the solution. A system with no source terms (free charges and currents) and media with finite conductivity will be assumed. Under these assumptions, the conductivity can be included in the complex permittivity.

In a linear and local medium with rotational invariance about the z-direction, the most general expressions for the tensor of the dielectric permittivity and for that of the magnetic permeability are

$$\begin{pmatrix} \varepsilon_\perp & 0 & 0 \\ 0 & \varepsilon_\perp & 0 \\ 0 & 0 & \varepsilon_z \end{pmatrix}, \qquad \begin{pmatrix} \mu_\perp & 0 & 0 \\ 0 & \mu_\perp & 0 \\ 0 & 0 & \mu_z \end{pmatrix}.$$

In any homogeneous region of this medium, the Maxwell equations can be written in terms of two independent components. Choosing the axial fields as independent fields, the curl equations can be expressed in cylindrical coordinates as [39]

$$\left( \frac{\partial^2}{\partial z^2} + \frac{\omega^2}{c^2} \varepsilon_\perp \mu_\perp \right) H_\rho = j \frac{\omega \varepsilon_\perp}{c} \frac{1}{\rho} \frac{\partial E_z}{\partial \varphi} + \frac{\partial^2 H_z}{\partial z \partial \rho},$$

$$\left( \frac{\partial^2}{\partial z^2} + \frac{\omega^2}{c^2} \varepsilon_\perp \mu_\perp \right) E_\rho = \frac{\partial^2 E_z}{\partial z \partial \rho} - j \frac{\omega \mu_\perp}{c} \frac{1}{\rho} \frac{\partial H_z}{\partial \varphi},$$

$$\left( \frac{\partial^2}{\partial z^2} + \frac{\omega^2}{c^2} \varepsilon_\perp \mu_\perp \right) H_\varphi = -j \frac{\omega \varepsilon_\perp}{c} \frac{\partial E_z}{\partial \rho} + \frac{1}{\rho} \frac{\partial^2 H_z}{\partial z \partial \varphi},$$

$$\left( \frac{\partial^2}{\partial z^2} + \frac{\omega^2}{c^2} \varepsilon_\perp \mu_\perp \right) E_\varphi = \frac{1}{\rho} \frac{\partial^2 E_z}{\partial z \partial \varphi} + j \frac{\omega \mu_\perp}{c} \frac{\partial H_z}{\partial \rho}.$$

In a similar fashion, the divergence equations yield

$$\varepsilon_\perp \frac{1}{\rho} \frac{\partial (\rho \cdot E_\rho)}{\partial \rho} + \varepsilon_\perp \frac{1}{\rho} \frac{\partial E_\varphi}{\partial \varphi} = -\varepsilon_z \frac{\partial E_z}{\partial z}$$

and

$$\mu_\perp \frac{1}{\rho} \frac{\partial (\rho \cdot H_\rho)}{\partial \rho} + \mu_\perp \frac{1}{\rho} \frac{\partial H_\varphi}{\partial \varphi} = -\mu_z \frac{\partial H_z}{\partial z}.$$

In the case of solutions with azimuthal invariance, the above equations decouple in two groups, the first one involving the field components ($H_\rho$, $E_\varphi$, $H_z$) and the second one the components ($E_\rho$, $H_\varphi$, $E_z$). The former admits therefore $TE_0$ solutions and the latter $TM_0$ solutions, which can be in both cases bounded or unbounded. The subscript '0' indicates the azimuthal invariance of the solution. Such two complete sets of modes do not share any field component. However, they can be mixed by the boundary conditions.

Consider now the boundary between two different regions labeled with 1 and 2, and the normal to the separating surface, labeled with $\hat{n}$. In the above assumptions, the boundary conditions reduce to continuity equations of the form $\begin{Bmatrix} \vec{D}_2 - \vec{D}_1 \\ \vec{B}_2 - \vec{B}_1 \end{Bmatrix} \cdot \hat{n} = 0$ and $\begin{Bmatrix} \vec{E}_2 - \vec{E}_1 \\ \vec{H}_2 - \vec{H}_1 \end{Bmatrix} \times \hat{n} = \vec{0}$. Where defined, the vector $\hat{n}$ cannot have azimuthal component in virtue of the rotational invariance of the system; consequently, $\hat{n} = n_\rho \hat{\rho} + n_z \hat{z}$ at any point. The scalar boundary conditions impose therefore the continuity of the term $\varepsilon_\perp E_\rho n_\rho + \varepsilon_z E_z n_z$ and of the term $\mu_\perp H_\rho n_\rho + \mu_z H_z n_z$, whereas the vector conditions impose the continuity of the components $E_\varphi$ and $H_\varphi$, and of the terms $n_\rho E_z - n_z E_\rho$ and $n_\rho H_z - n_z H_\rho$. Accordingly, also the boundary conditions can be separated in two sets of equations involving the above field components. Since these conditions do not mix the $TE_0$ and $TM_0$ partial solutions obtained in each homogenous region, the complete solutions resulting for the whole domain will show the same $TE_0$ or $TM_0$ character. The generalization of this result to systems with a continuous variation of permittivity and permeability can be obtained by approximating these quantities with stepwise functions, and then taking the limit. A limit operation on the conductivity of the employed materials can be employed as well to extend the above property to configurations including ideal conductors.

A similar approach can be followed to demonstrate that the modes with azimuthal and axial invariance ($\frac{\partial}{\partial \varphi} \equiv 0$, $\frac{\partial}{\partial z} \equiv 0$) propagating along a parallel-plate waveguide are, in the limit of ideal conductors, transverse electromagnetic (TEM) with nonvanishing components given by ($H_\varphi$, $E_z$).

# References

[1] P. P. Woskov, V.S. Bajaj, M. K. Hornstein, R. J. Temkin, R. G. Griffin, IEEE Trans. Microwave Theory Tech. **MTT-53,** 1863 ( 2005)
[2] D. R. Solli, C. F. McCormick, R. Y. Chiao, J. M. Hickmann, J. Appl. Phys. **93**, 9429 (2003)
[3] C. R. Boyd, IEEE Trans. Microwave Theory Tech. MTT-53, 2371 (2005)
[4] H. Blok, J. A. J. M. Disselhorst, H. van der Meer, S. B. Orlinskii, and J. Schmidt, J. Magn. Reson. **173**, 49 (2005)
[5] A. I. Meshkov, F. C. De Lucia, Rev. Sci. Instrum. **76**, 083103 (2005)
[6] E. M. Ganapolskii, A. V. Golik, A. P. Koroljuk, Phys. Rev. B **51**, 11962 (1995)
[7] F. M. Peeters, Superlattices and Microstructures **6**, 217 (1989)
[8] P. Exner, Phys. Lett. A **141**, 213 (1989)
[9] R. L. Schult, D. G. Ravenhall, and H. W. Wyld, Phys. Rev. B **39**, R5476 (1989)
[10] P. Exner, P. Seba, and P. Stovicek, Phys. Lett. A **150**, 179 (1990)
[11] Y. Avishai, D. Bessis, B. G. Giraud, and G. Mantica, Phys. Rev. B **44**, 8028 (1991)
[12] P. Exner and P. Seba, J. Math. Phys. **30**, 2574 (1989)
[13] P. Exner and P. Seba, Phys. Lett. A **144**, 347 (1990)
[14] J. Goldstone and R. L. Jaffe, Phys. Rev. B **45**, 14100 (1992)
[15] J. P. Carini, J. T. Londergan, K. Mullen, and D. P. Murdock, Phys. Rev. B **46**, 15538 (1992)
[16] J. P. Carini, J. T. Londergan, K. Mullen, and D. P. Murdock, Phys. Rev. B **48**, 4503 (1993)

[17] J. P. Carini, J. T. Londergan, D. P. Murdock, D. Trinkle, and C. S. Yung, Phys. Rev. B **55**, 9842 (1997)
[18] G. Annino, H. Yashiro, M. Cassettari, M. Martinelli, Phys. Rev. B **73**, 125308 (2006)
[19] G. Annino, M. Cassettari, M. Martinelli, Rev. Sci. Instrum. **76**, 084702 (2005)
[20] G. Annino, M. Cassettari and M. Martinelli, Rev. Sci. Instrum. **76**, 064702 (2005)
[21] G. Annino, M. Cassettari, M. Fittipaldi, M. Martinelli, J. Magn. Reson. **176**, 37 (2005)
[22] W. Lawson, J. Cheng, J. P. Calame, M. Castle, B. Hogan, V. L. Granatstein, M. Reiser, G. P. Saraph, Phys. Rev. Lett. **81**, 3030 (1998)
[23] Z. Zhai, C. Kusko, N. Hakim, S. Sridhar, A. Revcolevschi, A. Vietkine, Rev. Sci. Instrum. **71**, 3151 (2000)
[24] S. S. Chauhan, C. C. Rajyaguru, H. Ito, N. Yugami, Y. Nishida, T. Yoshida, Rev. Sci. Instrum. **72**, 4344 (2001)
[25] E. Gaganidze, R. Heidinger, J. Halbritter, A. Shevchun, M. Trunin, H. Schneidewind, J. Appl. Phys. **93**, 4049 (2003)
[26] J. Kim, K. Lee, B. Friedman, D. Cha, Appl. Phys. Lett. **83**, 1032 (2003)
[27] T. P. Crowley, E. A. Donley, T. P. Heavner, Rev. Sci. Instrum. **75**, 2575 (2004)
[28] M. Mola, S. Hill, P. Goy, and M. Gross, Rev. Sci. Instrum. **71**, 186 (2000); http://www.abmillimetre.com
[29] G. Annino, M. Cassettari and M. Martinelli, "Millimeter Waves Dielectric Resonators", in *Recent Research Developments in Microwave Theory and Techiques,* vol. 2, 195-236 (Transworld Research Network, Kerala (India), 2004)
[30] G. Annino, M. Cassettari, M. Fittipaldi, and M. Martinelli, J. Magn. Reson. **157**, 74 (2002)
[31] G. V. Stupakov, S. S. Kurennoy, Phys. Rev. E **49**, 794 (1994)
[32] S. S. Kurennoy, Phys. Rev. E **51**, 2498 (1995)
[33] L. Z. Duan, K. Gibble, Opt. Lett. **30**, 3317 (2005)
[34] A. Corney, *Atomic and laser spectroscopy*, (Oxford University Press, Oxford, 1988), chapt. 12 and 13
[35] C. P. Poole, *Electron Spin Resonance: a Comprehensive Treatise on Experimental Techniques* (Wiley, New York, 1983).
[36] S. M. Spillane, T. J. Kippenberg, K. J. Vahala, K. W. Goh, E. Wilcut, H. J. Kimble, Phys. Rev. A **71**, 013817 (2005)
[37] H. A. Atwater, *Introduction to Microwave Theory* (McGraw-Hill, New York, 1962)
[38] J. T. Robinson, C. Manolatou, L. Chen, M. Lipson, Phys. Rev. Lett. vol. 95, 143901 (2005)
[39] D. Kajfez and P. Guillon, *Dielectric Resonators* (Artech House, Dedham, 1986)